\documentclass{article}
\usepackage{hiph-art}
\volnumber{} \issuenumber{} \edyear{}                             %|
\frompage{} \topage{}                                              %|
\recrevdate{}                                              %|
\def\rightmark{Thermalization of gluons at RHIC}
\def\leftmark{Z. Xu and C. Greiner}
 
\title{Thermalization of gluons at RHIC: Dependence on initial conditions} 
\authors{ 
{Zhe Xu$^1$ and Carsten Greiner$^1$ %
\index{Xu, Z.} % Abbreviated names of the author(s),
\index{Greiner, C.} % to be inserted for use in the volume index
}\\[2.812mm]
{\normalsize
\hspace*{-8pt}$^1$ Institut f\"ur Theoretische Physik,
        Johann Wolfgang Goethe Universit\"at Frankfurt,
        Max-von-Laue Str. 1, D-60438 Frankfurt, Germany \\[0.2ex] 
}}
 
\abstract{
We investigate how thermalization of gluons depends on the initial conditions
assumed in ultrarelativistic heavy ion collisions at RHIC. The study
is based on simulations employing the pQCD inspired parton cascade solving
the Boltzmann equation for gluons. We consider independently produced
minijets with $p_T > p_0=1.3 \sim 2.0$ GeV and a color glass condensate as
possible initial conditions for the freed gluons. It turns out that
full kinetic equilibrium is achieved slightly sooner in denser system 
and its timescale tends to saturate. Compared with the kinetic equilibration
we find a stronger dependence of chemical equilibration on the initial
conditions.}


 
\makeindex
\begin{document}
 
\maketitle

\section{Introduction}\label{intro}

Recently we have studied kinetic and chemical equilibration of
gluons in a central heavy ion collision at RHIC energy employing a 
microscopical transport model including multiple $gg \leftrightarrow ggg$ 
scatterings \cite{xu05}. The results showed that even for an initially
dilute system, $dN_g/dy \approx 200$, when the initial conditions are
assumed as independent minijets with $p_T > p_0=2.0$ GeV, overall kinetic
equilibrium is achieved at $2$ fm/c and full chemical equilibirum follows
later at about $3$ fm/c. In addition, the expansion of the considered parton 
system behaves (quasi)hydrodynamically. Especially the total transverse energy
per unit rapidity at midrapidity, $dE_T/dy|_{y=0}$, decreases from initial
$500$ GeV to $270$ GeV at final time $4$ fm/c due to the longitudinal work
done by the pressure built up during the thermal equilibration.

Since the experiments at RHIC at $\sqrt{s}=200$ GeV obtained 
$dE_T/dy|_{y=0}=620\pm 33$ GeV \cite{STAR04} for the $5\%$ most central
events, our final value is lower by a factor of more than two when assuming 
that hadronization does not change the local energy density.
The cutoff $p_0=2$ GeV used in the simulations has been a rather 
conservative assumption for the initial minijets.
{\it Soft} partons with $p_T < p_0$, which are produced by the
nonperturbative part of nucleus-nucleus interaction and also
contribute to the entropy production, are completely neglected.
Such partons may stem from the color glass condensate
(CGC) \cite{mv94} and should also be included in the transport simulation,
when they are freed from the color field. 
In principle, one should combine gluons from CGC and harder
gluons from minijets production to give a more realistic initial
condition for partons after a heavy ion collision. The final results
after parton evolution with such initial condition is thus appropriate
for comparisons with the experimental data. We leave this interesting
topic for future investigations. For the moment, initial conditions
of gluons from CGC and minijets production are assumed separately for
independent simulations.

In another idea within a saturation picture of quarks and gluons in phase
space, presented by Eskola et al. \cite{EKRT00}, the authors assumed that
performing the computation of minijets at a saturation momentum, $p_0=p_{sat}$,
gives an estimate of the effect from all momentum scales, both
above and below $p_{sat}$. At RHIC energy they got $p_{sat}=1.13$ GeV
for corresponding initial minijets, and estimated the final transverse
energy $dE_T/dy|_{y=0} \approx 660$ GeV by assuming {\it immediate} 
thermalization and adiabatic expansion. Compared with the experimental
data, the calculations met the value of the transverse energy.
The smaller the cutoff $p_0$ is assumed, the larger will be the initial 
gluon density. Intuitively, a denser system may achieve faster 
thermalization than a dilute system. However, immediate thermalization 
might be a naive assumption, particularly for chemical equilibration.
This strongly motivates us to apply the developed parton cascade to 
inspect the timescale of thermalization in dependence on
the cutoff parameter $p_0$.

In addition, Serreau and Schiff \cite{ss01} have studied the role of elastic 
processes in kinetic equilibration using minijets as well as color glass
condensate as initial conditions and found that in all cases elastic 
interactions alone cannot drive the system rapidly to kinetic equilibrium.
It is interesting to see whether the situation changes when the inelastic
$gg \leftrightarrow ggg$ processes are included.

\section{Initial Conditions}

The production of the primary partons at the very onset of a 
central Au Au collision is assumed first as a free superposition of
minijets being liberated in the individual semihard nucleon-nucleon
interactions. We introduce an additional {\it formation time} for every 
minijet, $\Delta t_f=\cosh y \, \Delta \tau_f \approx \cosh y \cdot 1/p_T$,
which models the prior off-shell propagation of the to be freed,
on-shell partons in the scattering process.
Within that time span, one assumes, for simplicity, that the
still virtual parton does not interact and moves with speed of light.
The $p_0$ dependence of initial conditions is shown in Table \ref{tab1}. 
\begin{table}[htb] 
%\vspace*{-12pt}
\caption[]{Gluon number and transverse energy per unit rapidity at
midrapidity extracted from different initial conditions and at final
time $4$ fm/c after cascade simulations.}
\label{tab1}
\vspace*{-5pt}
\begin{center}
\begin{tabular}{lllll}
\hline\\[-10pt]
minijets/CGC & $\frac{dN_g}{dy}$(init.) & $\frac{dN_g}{dy}$($4$ fm/c) &
$\frac{dE_T}{dy}$(init.) & $\frac{dE_T}{dy}$($4$ fm/c) \\
$p_0$/$Q_s$ [GeV] & & & [GeV] & [GeV] \\ 
\hline\\[-10pt]
$p_0=2.0$ & 181 & 352 & 478 & 272 \\
$p_0=1.5$ & 537 & 663 & 1082 & 514 \\ 
$p_0=1.4$ & 688 & 781 & 1301 & 626 \\ 
$p_0=1.3$ & 889 & 930 & 1569 & 775 \\
$Q_s=1.0$ & 830 & 528 & 550 & 314 \\
\hline 
\end{tabular}
\end{center}
\end{table}
The calculations are performed by multiplying the number of individual
nucleon-nucleon collisions for a Au Au collision and the number of 
gluons produced in a nucleon-nucleon collision at $\sqrt{s}=200$ GeV.
The former is obtained by applying a Glauber picture with a Woods-Saxon
nuclear distribution. The latter is evaluated
employing the Gl\"uck-Reya-Vogt parametrizations \cite{GRV95} for
the parton structure functions. One sees that the initial gluon number
and transverse energy increase with decreasing $p_0$.

For the initial gluon distribution from the saturation picture we employ
an idealized and boost-invariant form proposed by Mueller \cite{M00} and
express it as
\begin{equation}
\label{fcgc}
f(x,p)|_{z=0}=\frac{c}{\alpha_s \, N_c}\, \frac{1}{\tau_f}\, \delta(p_z)\,
\Theta(Q_s^2-p_T^2)\,,
\end{equation}
which is described by a bulk scale $Q_s$, the momentum scale at
which gluon distribution saturates. The boost-invariance leads to
the equality of momentum and space-time rapidity, i.e., $\eta=y$,
for the initial gluons. Distribution (\ref{fcgc}) has been
used in \cite{BV01} to investigate kinetic equilibration of gluons 
in the subsequent evolution by means of a non-linear Landau equation. 
In contrast to the minijets the saturation model gives the distribution 
of partons with $p_T < Q_s$. We read off the value of parameters taken 
in \cite{BV01}: $N_c=3$ for SU(3), $c=1.3$, $\alpha_s=0.3$, $Q_s=1$ GeV
and the corresponding formation time, at which gluons become on-shell,
to be $\tau_f=0.4$ fm/c $\sim 1/Q_s$, and assume that gluons will be 
produced within a transverse radius of $6$ fm (Au nucleus) and within
$|\eta| < 3$ longitudinally. Then
we obtain $dN_g/dy|_{y=0}=830$ and $dE_T/dy|_{y=0}=550$ GeV, which are 
also included in Table \ref{tab1} for comparisons.

\section{Three-body Gluonic Interactions}

The three-body gluonic interactions are described by the matrix element
\cite{biro93}
\begin{equation}
\label{m23}
| {\cal M}_{gg \to ggg} |^2 = \left ( \frac{9 g^4}{2} 
\frac{s^2}{({\bf q}_{\perp}^2+m_D^2)^2} \right ) 
\left ( \frac{12 g^2 {\bf q}_{\perp}^2}
{{\bf k}_{\perp}^2 [({\bf k}_{\perp}-{\bf q}_{\perp})^2+m_D^2]} \right )
\Theta(k_{\perp}\Lambda_g-\cosh y) \, ,
\end{equation} 
where $g^2=4\pi\alpha_s$. ${\bf q}_{\perp}$ and ${\bf k}_{\perp}$ denote,
respectively, the perpendicular component of the momentum transfer and 
that of the momentum of the radiated gluon in the c.m. frame of the 
collision. We regularize the infrared divergences by using the Debye
screening mass $m_D^2$ which is calculated locally over the present
particle density obtained from the simulation. The suppression of
the radiation of soft gluons due to the Landau-Pomeranchuk-Migdal (LPM)
effect, which is expressed via the step function in Eq. (\ref{m23}),
is modeled by the consideration that the time of the emission,
$\sim \frac{1}{k_{\perp}} \cosh y$, should be smaller than the time
interval between two scatterings or equivalently  the gluon mean free
path $\Lambda_g$. This leads to a lower cutoff of $k_{\perp}$ and
to a decrease of the total cross section. The dependence of
the total cross section on the LPM-cutoff is approximately logarithmic
and thus its sensitivity to the dynamics is only moderate.

\section{Results}

The space time propagations of gluons are simulated until a final time
$4$ fm/c at which the energy density at midrapidity at $p_0=2.0$ GeV has
already decreased nearly to the critical value of $1$ $\mbox{GeV/fm}^3$ being 
characteristic for hadronization. The gluon number and transverse energy
per rapidity at the final time are shown in Table \ref{tab1}. Comparing with
the initial values the enhancement of gluon number is weaker at smaller
$p_0$, which implies larger initial fugacity at smaller $p_0$. 
On the contrary, the initial gluons from the saturation model are, 
surprisingly, {\it oversaturated}. This can be interpreted as follows. 
According to the gluon distribution (\ref{fcgc}) the initial effective 
temperature is $T(\tau_f)=\epsilon/3n=2/9*Q_s$. We obtain then
the effective gluon fugacity at the formation time
\begin{equation}
\lambda_g(\tau_f)=\frac{n}{n_{eq}}=\frac{\frac{dN}{\pi R^2\tau_f d\eta}}
{2(N_c^2-1) \frac{T^3}{\pi^2}}=\frac{9^3}{4^3}\,\frac{c}
{\alpha_s N_c \tau_f Q_s}\,,
\end{equation}
which would be valid when assuming immediate kinetic equilibrium at
$\tau=\tau_f$. For the setup of parameters given before one has 
$\lambda_g(\tau_f)=8$ ! So after the gluons are freed from the color 
field and become on-shell, $ggg\to gg$ collisions will be the dominant
processes driving the system into chemical equilibrium.

\begin{figure}[htb]
\vspace*{0.2cm}
\begin{center}
\epsfysize=5cm 
\epsfbox{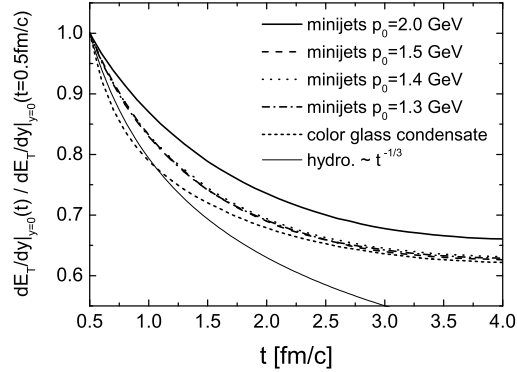}
\end{center}
\vspace*{-0.4cm}
\caption{Time evolution of nomalized transverse energy per unit rapidity
at midrapidity. Results are obtained from simulations with
different initial conditions of gluons, assumed in a central Au Au
collision at RHIC energy.}
%\vspace*{-0.2cm}
\label{et}
\end{figure}

\begin{figure}[htb]
%\vspace*{0.2cm}
\begin{center}
\epsfysize=6cm 
\epsfbox{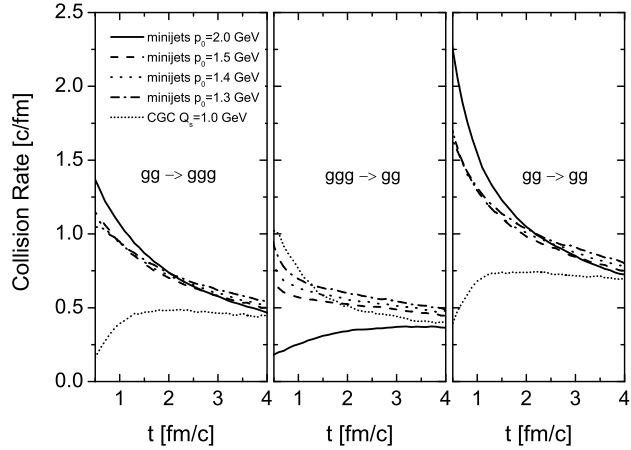}
\end{center}
\vspace*{-0.4cm}
\caption{Time evolution of collision rate extracted in the central region
taken as an expanding cylinder with a radius of $1.5$ fm and within 
a unit interval of space time rapidity $\eta$ around the collision 
center $\eta=0$.}
%\vspace*{-0.2cm}
\label{rate}
\end{figure}

From Table \ref{tab1} one also sees that in all cases of initial
conditions the transverse energy per unit rapidity decreases with
the progressing time due to the onset of collective expansion. 
The final values of $dE_T/dy|_{y=0}$ are, however, quite different. 
We see that the value at $p_0=1.4$ GeV would meet the experimental data.
Thus the minijets production at $p_0=1.4$ GeV seems to give an appropriate
initial condition of gluons at RHIC. To show the decreasing of the 
transverse energy at midrapidity we depict in Fig. \ref{et} the
normalized transverse energies obtained from simulations with the
different initial conditions.
The denominators are the transverse energies at midrapidity at time
$0.5$ fm/c, which are nearly the initial values shown in Table \ref{tab1}.
We see a unique behavior of the time evolutions of the normalized
$dE_T/dy|_{y=0}$, especially for initial conditions employing minijets 
with $p_0=1.3-1.5$ GeV and CGC with $Q_s=1.0$ GeV. For these initial 
conditions the decreasing of the transverse energy is scaled with the
initial value. This indicates that the work done by the pressure
is also scaled with the initial pressure when it is built up due to
thermalization. When looking at the time evolution of collision rate
per particle, depicted in Fig. \ref{rate} for all scattering channels, 
we see that the rates are in the same range at late times.
This seems to be the reason for the scaling behavior observed for
$dE_T/dy|_{y=0}(t)$, since the total collision number per unit time, 
which corresponds to the work done by the pressure, is proportional to
the present local gluon density which is not very different from
the initial value for the cases at $p_0=1.3-1.5$ GeV, seen in
Table \ref{tab1}. On the other hand one should realize that small angle
scatterings do not really contribute much to (quasi)hydrodynamical work.
{\it Transport} cross sections, or better, {\it transport} mean free path
should and will be discussed in detail within the parton transport
approach in future investigations.

In Fig. \ref{et} we also depict the normalized transverse energy per
unit rapidity for an ideal hydrodynamical expansion by the thin solid
line. Comparisons show that the collective expansion due to the pQCD 
interactions is quasi ideal at $0.5-2$ fm/c. At late times the decrease
of $dE_T/dy|_{y=0}$ slows down, since the collision rates become smaller,
especially in the outer, transversally expanding
region where the gluon density is smaller compared to the central region.

Inspecting Fig. \ref{rate} again and focusing on the cases at 
$p_0=1.3-1.5$ GeV one finds that, at a certain time, the smaller $p_0$ is, 
the larger are the rates. This is consistent with the finding shown in Fig. 11
in \cite{xu05} that at equilibirum the collision rate is proportional 
to temperature $T$ which is larger in denser system initialized at 
smaller $p_0$. The cross section is then approximately inverse 
proportional to $T^2$. Therefore, at a certain time, the cross 
sections $\sigma_{gg\to gg}$ for $gg \to gg$ and $\sigma_{gg\to ggg}$
for $gg \to ggg$ scatterings are smaller at smaller $p_0$. For instance,
comparing with the time evolution of cross sections calculated
at $p_0=2.0$ GeV, $\sigma_{gg\to gg}$($\sigma_{gg\to ggg}$)
obtained from the simulation at $p_0=1.4$ GeV increases from $0.5(0.3)$ mb
at $0.5$ fm/c to $2.5(1.4)$ mb at $4$ fm/c respectively. The values at
the final time are approximately $1.6$ times smaller than those at 
$p_0=2.0$ GeV, which is consistent with the ratio of $T^2$ at $4$ fm/c
[$T=215(262)$ MeV at $p_0=2.0(1.4)$ GeV].

\begin{figure}[htb]
%\vspace*{0.2cm}
\epsfysize=4.5cm 
\epsfbox{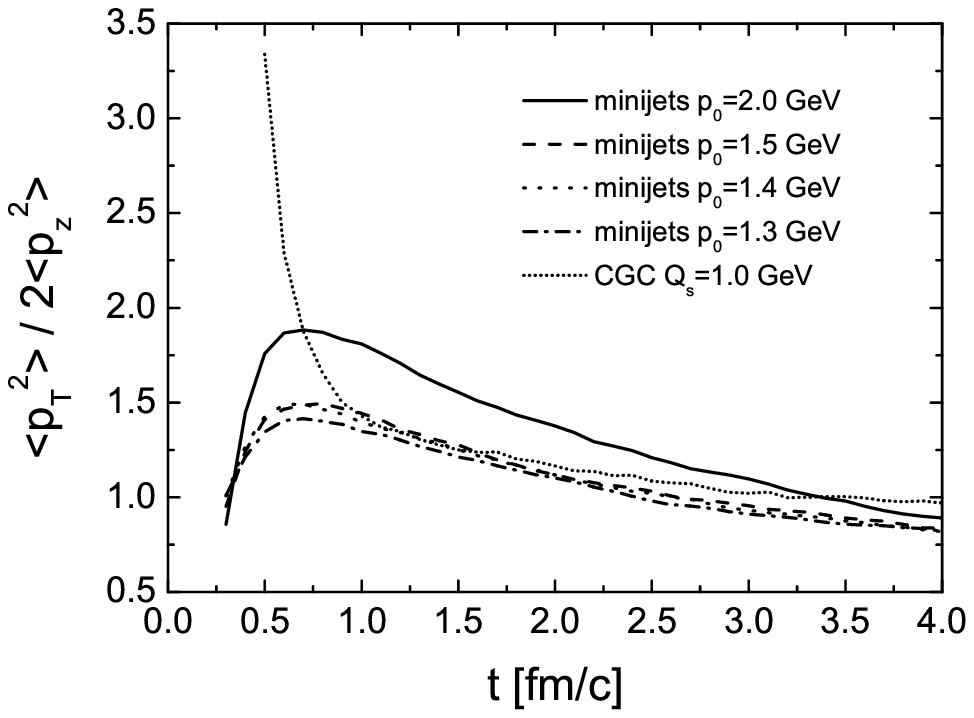}
\hspace{\fill}
\epsfysize=4.5cm 
\epsfbox{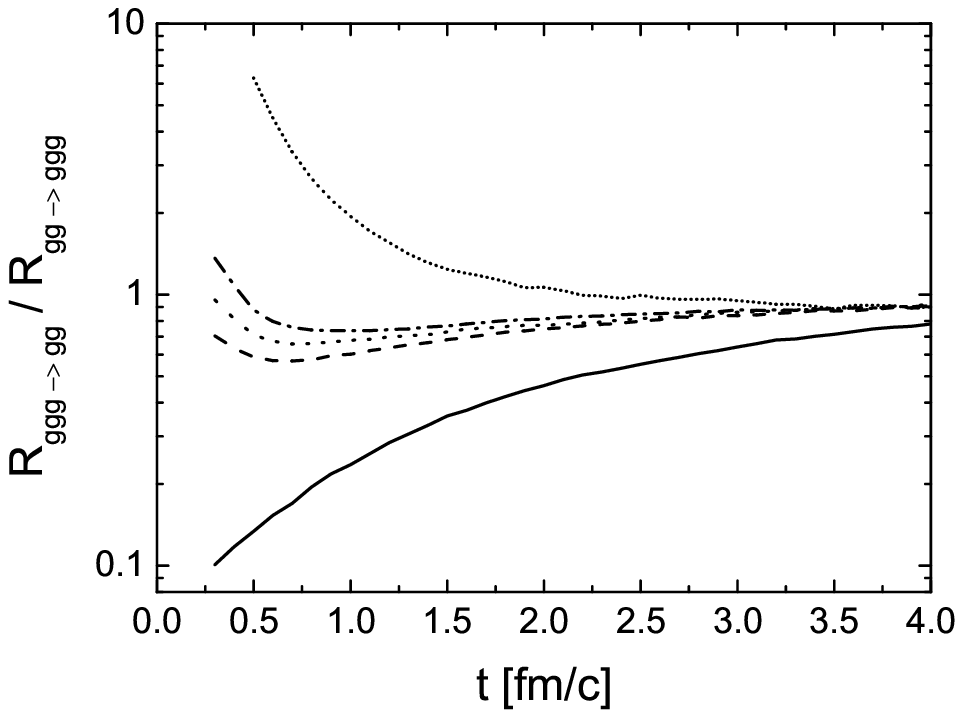}
\vspace*{-0.2cm}
\caption{Left panel: Time evolution of the momentum anisotropy.
Right panel: Time evolution of the effective fugacity
$\lambda_g=R_{ggg\to gg}/R_{gg\to ggg}$. Results are 
extracted in the central region of the expanding gluon systems simulated
using the parton cascade with different initial conditions.}
%\vspace*{-0.2cm}
\label{therm}
\end{figure}

Figure \ref{therm} shows kinetic and chemical equilibration observed
locally in the central region.
The kinetic equilibration is characterized by the time evolution of
the momentum anisotropy, $<p_T^2>/2<p_z^2>$ , which maintains a value
of one at equilibrium. For an extreme case of free streaming the
final momentum distribution of particles in a local disc would be 
transversely directed, which leads to an infinite value of the momentum
anisotropy. Therefore initial free streaming is the reason for the 
increase of the momentum anisotropy at the beginning of the expansion,
seen in Fig. \ref{therm} from the simulations with minijets initial
conditions. The following bending over signals the onset of equilibration.
We also see that full kinetic equilibrium comes slightly sooner, at
$1-2$ fm/c, at smaller $p_0$. The timescale tends to saturate at smaller
$p_0$. For the case using CGC as initial condition, the momenta of
gluons are initially directed transversely in local discs according to
Eq. (\ref{fcgc}). Therefore the momentum anisotropy has a large initial
value. Surprisingly, from $0.8$ fm/c the time evolution of the
momentum anisotropy is almost identical with that obtained from simulations
at smaller $p_0$.

\begin{figure}[htb]
\vspace*{-0.1cm}
\begin{center}
\epsfysize=5cm 
\epsfbox{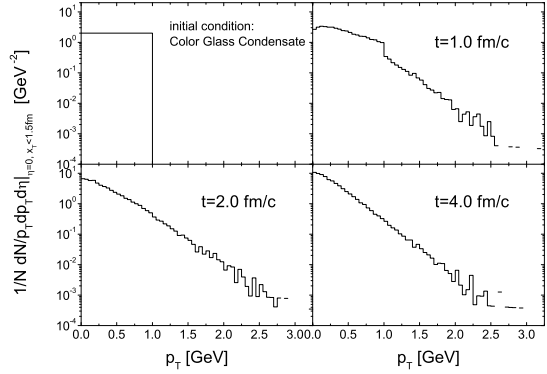}
\end{center}
\vspace*{-0.5cm}
\caption{Transverse momentum spectrum in the central region.}
\vspace*{-0.1cm}
\label{ptcgc}
\end{figure}

However, we note that the maintenance of the momentum isotropy 
is not necessaryly equivalent to kinetic equilibrium, because even a
free streaming keeps the momentum isotropy, but leads the system out
of equilibrium. Therefore, one has to look at the behavior of 
the $p_T$ spectrum in time. While a free streaming can not change
the $p_T$ spectrum, the spectrum steepens continuously when the system
maintains equilibrium, since the thermodynamical work outwards cools
down the system. When we compare the $p_T$ spectra
at $2.0$ and $4.0$ fm/c, depicted in Fig. \ref{ptcgc} for the case when 
a color glass condensate is assumed as the initial condition of gluons,
we realize that the local system at the central region is indeed at 
equilibrium at late times. One finds the same result also for minijets 
initial conditions as shown in more detail in Fig. 44 in 
reference \cite{xu05}. At late times, for the transversely outer region 
the situation is different. The local system becomes out of kinetic 
equilibrium when one goes transversely away from the center to the edge
of the system. That is the reason why the total transverse energy at 
midrapidity decreases only moderately at late times, as shown in Fig. \ref{et}.

In the right panel of Fig. \ref{therm} we show the ratio of the
collision rates for $ggg \to gg$ and $gg \to ggg$ scatterings, which 
is equal to the gluon fugacity $\lambda_g$ when the system is at kinetic 
equilibrium. The collision rates have been already shown in Fig. \ref{rate}.
We realize that the timescale for full chemical equilibrium depends
on the initial conditions. At $p_0=2.0$ GeV the system is initially
undersaturated and reaches full equilibrium at $3$ fm/c. At smaller
$p_0$ the system becomes denser and chemical equilibrium is achieved
almost initially. For a CGC situation the initial system is oversaturated. Full
chemical equilibrium follows at a time $\approx 1.5$ fm/c.

Finally, looking at the transverse momentum spectrum depicted in 
Fig. \ref{ptcgc},
we observe that the region around the hard scale $Q_s=1.0$ GeV, which
is mostly populated by initial gluons, is less and less occupied with 
progressing time. The dominantly populated region moves to smaller $p_T$
scale as cooling proceeds. At $2$ fm/c the spectrum possesses a full 
thermal shape. Thermalization somehow resembles the idealistic 
bottom-up scenario \cite{baier01}. However, in clear contrast to that
scenario, the gluon number decreases for the particular parameter
sets of the initial condition.

\section{Conclusion}
We have studied local thermalization of gluons in a central Au Au
collision at RHIC energy. Initial conditions of gluons are assumed
by minijets production at different cutoff $p_0=1.3-2.0$ GeV and 
by a color glass condensate with the saturation scale $Q_s=1.0$ GeV
respectively. We apply the newly developed parton cascade including
$gg\leftrightarrow ggg$ to simulate the space time evolution of gluons.
The numerical results show that kinetic equilibrium comes sooner in
denser system and the timescale tends to saturate, while the dependence
of the chemical equilibration on the initial conditions is much stronger.

Future investigations will concentrate on the energy loss of 
high-energy partons, parametrical dependence of the timescale of 
thermalization on $Q_s$ and $\alpha_s$ concerning the bottom-up picture,
generation of elliptic flow $v_2$ in noncentral collisions and viscosity
in the plasma. Moreover, quarks will be included into the parton cascade.
A special interest will be put on the study of elliptic flow of heavy quarks.

\vfill\eject
\end{document}